\documentclass[rnote]{aa} 
\usepackage{natbib}
\bibpunct{(}{)}{;}{a}{}{,}

\usepackage{graphicx}
\usepackage{txfonts}
\begin{document}
   \title{Stellar Core Collapse with Hadron-Quark Phase Transition}


   \author{Ken'ichiro Nakazato\inst{1} \and Kohsuke Sumiyoshi\inst{2} \and Shoichi Yamada\inst{3,4}}

   \institute{Department of Physics, Faculty of Science \& Technology, Tokyo University of Science, 2641 Yamazaki, Noda, Chiba 278-8510, Japan
           \email{nakazato@rs.tus.ac.jp}
           \and
           Numazu College of Technology, Ooka 3600, Numazu, Shizuoka 410-8501, Japan
           \and
           Department of Physics, Faculty of Science \& Engineering, Waseda University, 3-4-1 Okubo, Shinjuku, Tokyo 169-8555, Japan
           \and
           Advanced Research Institute for Science \& Engineering, Waseda University, 3-4-1 Okubo, Shinjuku, Tokyo 169-8555, Japan
   }

   \date{Received \today}
 
  \abstract
   {Hadronic matter undergoes a deconfinement transition to quark matter at high temperature and/or high density. It would be realized in collapsing cores of massive stars.}
   {In the framework of MIT bag model, the ambiguities of the interaction are encapsulated in the bag constant. Some progenitor stars that invoke the core collapses explode as supernovae, and other ones become black holes. The fates of core collapses are investigated for various cases.}
   {Equations of state including the hadron-quark phase transition are constructed for the cases of the bag constant $B=90$, 150 and 250~MeV~fm$^{-3}$. To describe the mixed phase, the Gibbs condition is used. Adopting the equations of state with different bag constants, the core collapse simulations are performed for the progenitor models with 15 and $40M_{\odot}$.}
   {If the bag constant is small as $B=90$~MeV~fm$^{-3}$, an interval between the bounce and black hole formation is shortened drastically for the model with $40M_{\odot}$ and the second bounce revives the shock wave leading to explosion for the model with $15M_{\odot}$.}
   {}

   \keywords{black hole physics --- dense matter --- equation of state --- hydrodynamics --- methods: numerical --- supernovae:general}

   \maketitle
%

\section{Introduction}

Massive stars with the main-sequence mass $M \gtrsim 10M_\odot$ are known to cause core collapse at the end of their lives, leading to supernova explosions or black hole formations \citep[e.g.,][]{nomoto06}. During the collapse, some sort of phase transition such as a quark deconfinement, hyperon appearance or meson condensation may occur and affect the dynamics \citep[e.g.,][]{takahara85}. In this paper, we investigate the hadron-quark phase transition. Recently, \citet{sage09} and \citet{fish11} found that, if the critical density for the transition is enough low, the proto--neutron star collapses to a more compact quark star in hundreds of milliseconds after the first bounce. The second collapse and the following bounce leads to the formation of shock wave and successful supernova explosion. On the other hand, according to our previous studies \citep[][]{self08,self10}, the second collapse of very massive stars to black holes is triggered by the transition even if the critical density is not so low. In this study, we extend these works to other cases.

\section{Equation of State and Numerical Setups}

The hadron-quark phase transition affects the dynamics of core collapses through the equation of state (EOS). To construct the EOS, we follow the methods in \citet{self08}. Firstly, we prepare two types of EOSs, hadronic EOS and quark EOS. In this study, we utilize a table constructed by \citet{shen98a,shen98b} based on relativistic mean field theory for the hadronic EOS. The MIT bag model of the deconfined 3-flavor strange quark matter is used for the quark EOS \citep[][]{bag74}. Note that the both EOSs include the effects of finite temperature, which is important to deal with the stellar core collapse. As a next step, we describe the phase transition and EOS for the mixed phase assuming the Gibbs conditions between the hadronic and quark EOSs \citep[][]{glende92}.

In the MIT bag model, free quarks are confined in the ``bag'' with a positive potential energy per unit volume, $B$. This parameter is called the bag constant and characterizes the EOS. For instance, the models with larger value of $B$ have higher transition density and temperature. It is known that, in the high temperature regime, the transition line has an end point called the critical point. Although \citet{self08} have shown that the EOS with $B=250$~MeV~fm$^{-3}$ (in another unit, $B^{1/4}=209$~MeV) can reproduce the temperature of critical point ($T_c \sim 170$~MeV), the bag ``constant'' may not be a constant. Namely, the bag ``constant'' for such a high temperature regime may effectively differ from that for the temperatures of our interest (tens of MeV). Therefore, we examine the cases with $B=90$~MeV~fm$^{-3}$, 150~MeV~fm$^{-3}$, and 250~MeV~fm$^{-3}$ ($B^{1/4}=162$~MeV, 184~MeV, and 209~MeV, respectively) for a systematic study. In particular, the value of $B^{1/4}=162$~MeV is adopted in \citet{sage09} and \citet{fish11}.

\begin{figure}
\centering
\includegraphics{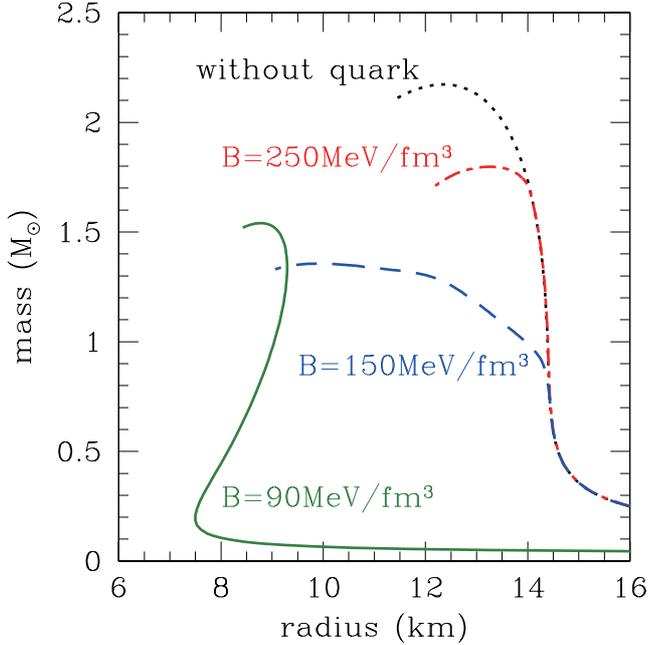}
\caption{Mass-radius relations of the compact stars for the cases with $B=90$~MeV~fm$^{-3}$ (solid), 150~MeV~fm$^{-3}$ (dashed), and 250~MeV~fm$^{-3}$ (dot-dashed) and for the case without quark transition (dotted).}\label{rmass}%
\end{figure}

We show the mass-radius relation of compact stars for our EOSs in Figure~\ref{rmass}. The maximum masses for the cases with $B=90$, 150, and 250~MeV~fm$^{-3}$ are 1.54, 1.36, and $1.80M_{\odot}$, respectively, while that for the case without quarks \citep[pure hadronic EOS,][]{shen98a,shen98b} is $2.17M_{\odot}$. The trajectories of $B=150$ and 250~MeV~fm$^{-3}$ align with that of pure hadronic EOS for low-mass stars. In these cases, even for the maximum-mass stars, quark matter at the central region is surrounded by hadronic matter inside the core. Stars with such structure are called ``hybrid'' stars. On the other hand, the trajectory of $B=90$~MeV~fm$^{-3}$ is quite different from others. Stars in this sequence are almost pure ``quark'' stars while thin hadronic crust is accompanied. Because of the ``transition'' from the hybrid stars to the quark stars, the maximum mass is not monotonic with respect to $B$. Note that, recently, the mass of the binary millisecond pulsar J1614-2230 was evaluated as $1.97 \pm 0.04M_\odot$ from a strong Shapiro delay signature \citep[][]{demo10}, and our models with quark cannot account for it. In order to support such larger masses, much larger bag constant as $B \gtrsim 400$~MeV~fm$^{-3}$ is needed in our framework \citep[][]{self08}. According to \citet{sage12}, the inclusion of the strong coupling constant with large value ($\alpha_s=0.7$) can increase the maximum mass of EOS with $B \sim 50$~MeV~fm$^{-3}$ up to $\ge 2M_{\odot}$. Alternatively, since the bag model dealt with here is a basic model for the quark-gluon-plasma, we may need to consider other phases for zero-temperature such as a color-flavor-locking or color superconductors \citep[e.g.,][]{fuku11}.

Using EOSs described above, we perform the core-collapse simulation as in \citet{self10}. We adopt the general relativistic implicit Lagrangian hydrodynamics code, which simultaneously solves the neutrino Boltzmann equations under spherical symmetry \citep[][]{yamada97,yamada99,sumi05}. For the hadronic phase, we follow the neutrino distribution functions from the reactions and transports. On the other hand, for the mixed phase and quark phase, we set them to be Fermi-Dirac functions because our EOS is constructed assuming that neutrinos are fully trapped and in equilibrium with other particles. As for the cases of black hole formation, we judge it by the appearance of the apparent horizon \citep[][]{self06}, which is a sufficient condition for the existence of the event horizon. Results of evolutionary calculations for progenitor stars with $M=15M_{\odot}$ and $40M_{\odot}$ by \citet{woosley95} are chosen as the initial models of our simulations.

\section{Collapse of $40M_{\odot}$ Star}

\begin{figure}
\centering
\includegraphics{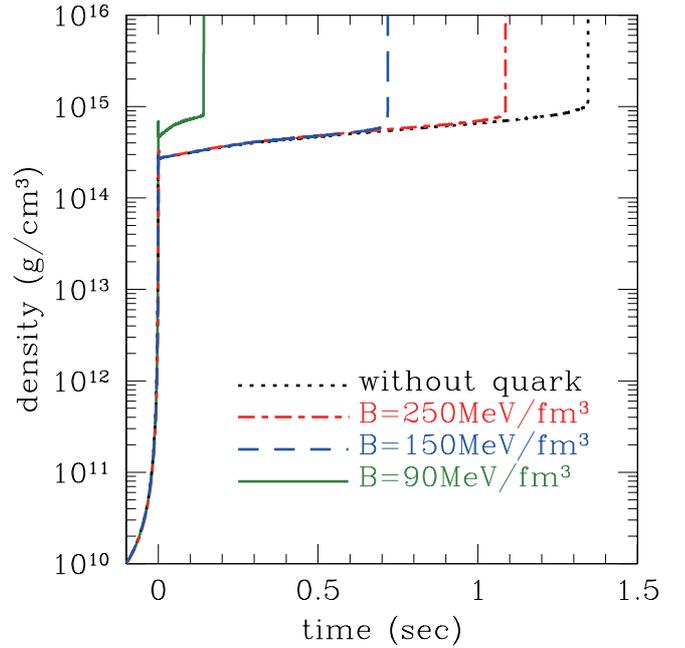}
\caption{Time evolutions of the central baryon mass density for the collapse of $40M_{\odot}$ star. The notation of lines is same as in Figure~\ref{rmass}.}\label{central}%
\end{figure}

\begin{figure*}
\centering
\includegraphics[scale=0.89]{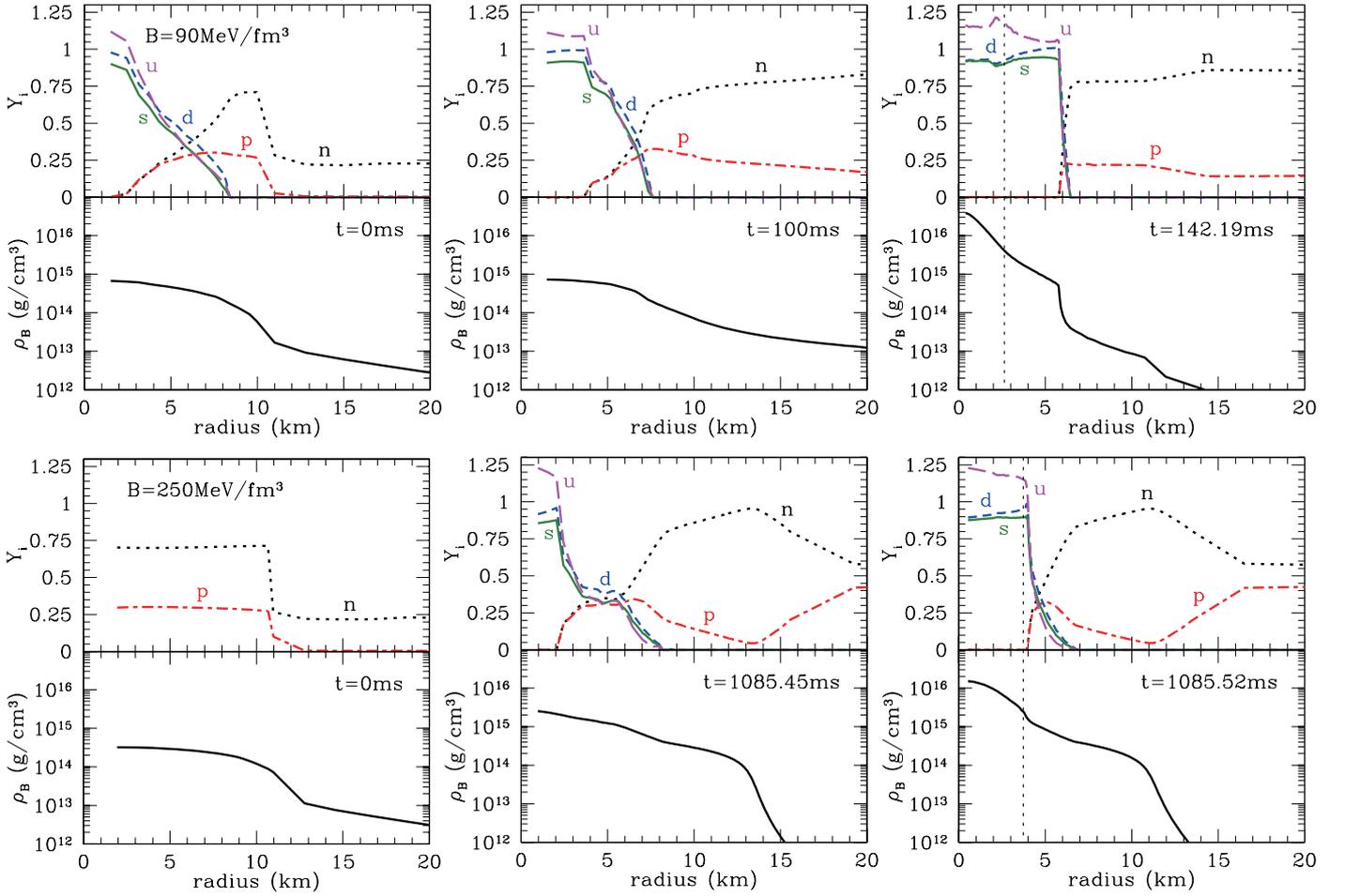}
\caption{Profiles of the particle fractions $Y_i \equiv n_i/n_B$ (upper plots) and baryon mass density $\rho_B$ (lower plots) for the collapse of $40M_{\odot}$ star, where $n_i$ represents the number density of particle $i$, and $n_B$ represents the baryon mass number density. For the upper plots of all panels, dotted, dot-dashed, long-dashed, short-dashed and solid lines correspond to neutron, proton, $u$ quark, $d$ quark and $s$ quark, respectively. Upper panels are for the case with $B=90$~MeV~fm$^{-3}$, and left, central, and right ones, respectively, show the profiles at 0~ms, 100~ms, and 142.19~ms after the bounce. Lower panels are for the case with $B=250$~MeV~fm$^{-3}$, and left, central, and right ones, respectively, show the profiles at 0~ms, 1085.45~ms, and 1085.52~ms after the bounce. Note that, the upper right and lower right panels correspond to the moment of black hole formation for each case, and thin dotted vertical lines show the location of the apparent horizon.}\label{comp}%
\end{figure*}

In the previous study \citep[][]{sumi07}, the collapse of $40M_{\odot}$ star adopted here was confirmed to result in the black hole formation. While this conclusion holds for the models with quark, the dynamics is affected. We show the time profiles of the central baryon mass density in Figure~\ref{central}. The bounce due to the nuclear force corresponds to the spikes at $t=0$, which is defined as the time of the bounce. The density at the bounce of the model of $B=90$~MeV~fm$^{-3}$ is $6.9\times10^{14}$~g~cm$^{-3}$ while those of other models are $3.2\times10^{14}$~g~cm$^{-3}$. In Figure~\ref{comp}, we show the profiles of the particle fractions and the baryon mass density of the models of $B=90$ and 250~MeV~fm$^{-3}$. As seen in this figure, deconfined quarks appear even at the bounce for the model of $B=90$~MeV~fm$^{-3}$. Thus the EOS gets softer and the bounce density becomes higher. On the other hand, the phase transition does not occur for the models of $B=150$ and 250~MeV~fm$^{-3}$ and their bounce densities do not differ from that of the model without quarks.

After the bounce, a shock wave is launched outward and a proto--neutron star is formed at the center. The shock wave stalls on the way and does not go out from the core. The proto--neutron star contracts gradually due to the accretion of shocked matter and, finally, collapses to a black hole, which corresponds to the blow-ups of the central density in Figure~\ref{central}. For the cases with $B=150$ and 250~MeV~fm$^{-3}$, quarks do not appear until just before the black hole formation and the transition triggers the second collapse. These features have already been reported in \citet{self10}. On the other hand, for the case with $B=90$~MeV~fm$^{-3}$, we find that quarks exist even at the early stage. Shortening of the interval between the bounce and black hole formation is drastic in this case.

The intervals between the bounce and black hole formation are 142.19~ms, 717.83~ms and 1085.52~ms for the models with $B=90$, 150 and 250~MeV~fm$^{-3}$, respectively, while it is 1345.40~ms for the model without quarks. This order is different from the order for the maximum masses of ``cold'' neutron stars. The proto--neutron star is not zero temperature and the maximum mass is increased by the thermal pressure. For the case with $B=90$~MeV~fm$^{-3}$, the proto--neutron star is compressed adiabatically and quark matter region gets larger. If the hadron-quark phase transition occurs adiabatically, the temperature decreases. This is because we assume that the transition is a first-order where the release of latent heat occurs \citep[][]{self10}. Note that, in this case, the low-density (hadron) phase has the lower entropy, which is opposite to an ordinary liquid-vapor transition (e.g., water vapor has lower density and higher entropy than liquid water in the liquid-vapor transition of H$_2$O). As a result, the thermal pressure is insufficient due to the temperature decrease and the proto--neutron star collapses quickly for the case with $B=90$~MeV~fm$^{-3}$. On the other hand, for the cases with $B=150$ and 250~MeV~fm$^{-3}$, quarks do not appear until just before the black hole formation and the maximum mass is sufficiently increased by the thermal pressure. 

According to \citet{self10}, the hadron-quark phase transition only affects the duration of neutrino emission and does not change the neutrino luminosity and spectrum before the black hole formation for the case with $B=250$~MeV~fm$^{-3}$. This is true even for the case with $B=90$~MeV~fm$^{-3}$. While the phase transition occurs inside the core, neutrinos are emitted from outer region. Therefore, the phase transition is not reflected in the neutrino signal very much within the timescale of the black hole formation for this model.

\section{Collapse of $15M_{\odot}$ Star}

In the previous study \citep[][]{sumi05}, it was shown that $15M_{\odot}$ star adopted here fails to explode for the spherical case without the hadron-quark phase transition, as is the case for almost other spherical simulations. This result would hold for the cases with $B=150$ and 250~MeV~fm$^{-3}$ because the density and temperature are not too high to occur the phase transition at least up to 1000~ms after the bounce \citep[][]{sumi05}. On the other hand, for the case with $B=90$~MeV~fm$^{-3}$, \citet{sage09} and \citet{fish11} found that the phase transition causes the second collapse and bounce leading to the successful supernova explosion with different progenitor models. Here, we verify this scenario for the model with $15M_{\odot}$ by \citet{woosley95} for the first time.

\begin{figure}
\centering
\includegraphics{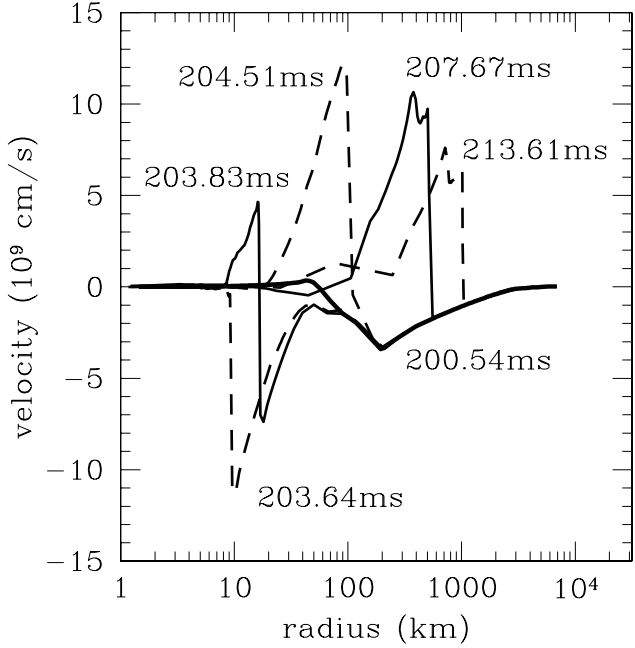}
\caption{Snapshots of velocity profile of the model with $15M_{\odot}$ for the case with $B=90$~MeV~fm$^{-3}$. Thick solid line corresponds to the onset of the second collapse (200.54~ms after the first bounce) and the second bounce occurs at 203.64~ms after the first bounce.}\label{vel}%
\end{figure}

We find that the second collapse starts at 200.54~ms after the first bounce and the second bounce occurs at 203.64~ms after the first bounce. The shock wave formed by the second bounce reaches at a radius of 1000~km within 10~ms, and the explosion is successful. In Figure~\ref{vel}, the snapshots of velocity profile for this stage are shown. We estimate the baryonic mass of remnant neutron star $M_{b,{\rm NS}}$ and supernova explosion energy $E_{\rm exp}$ with the data at 213.61~ms after the first bounce. First of all, we calculate the sum of kinetic energy and gravitational potential, $e_{k+g}$, for each mass element in the shocked region. We assume that the elements with $e_{k+g}<0$ fall back to the central neutron star. Then we get $M_{b,{\rm NS}}=1.42M_{\odot}$. As a next step, we integrate $e_{k+g}$ and the internal energy over the region with $e_{k+g}>0$ and a positive radial velocity. Adding the gravitational potential of outer pre-shocked region, we estimate as $E_{\rm exp}=1.54\times10^{51}$~erg. This is the same order of magnitude as the canonical value. Note that, a part of internal energy would be carried by the neutrino radiation. If we integrate only $e_{k+g}$, we get $E_{\rm exp}=1.7\times10^{50}$~erg. The real value of the explosion energy should reside between two estimations and this result is consistent with \citet{fish11}.

\begin{figure*}
\centering
\includegraphics[scale=0.89]{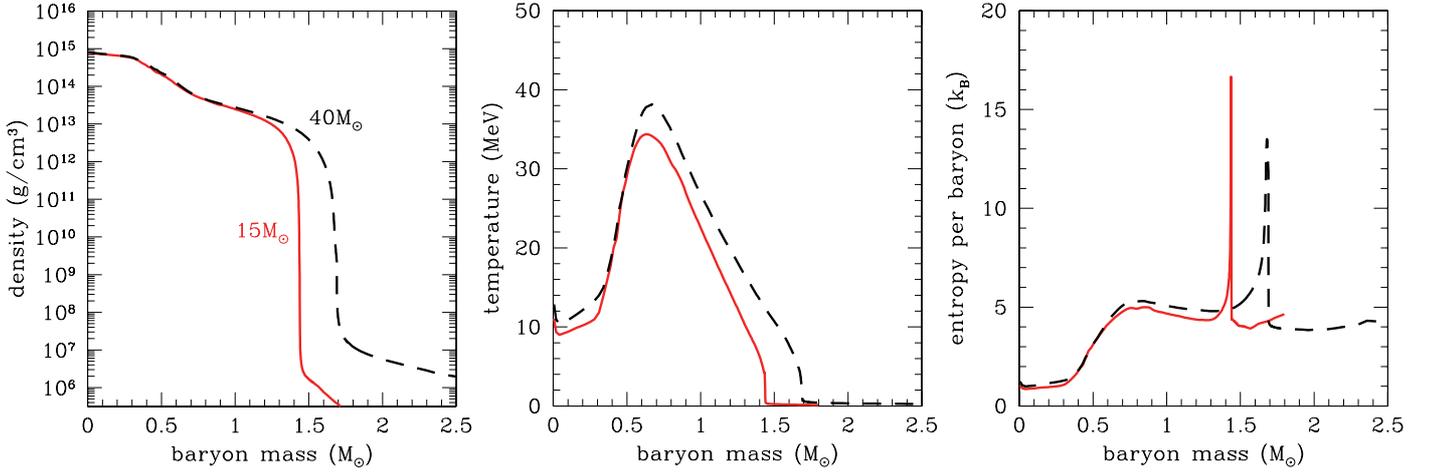}
\caption{Profiles of density (left), temperature (center) and entropy per baryon (right) for the case with $B=90$~MeV~fm$^{-3}$. Solid lines are for the model with $15M_{\odot}$ at 200.54~ms after the first bounce and dashed lines are for the model with $40M_{\odot}$ at 139.23~ms after the first bounce. Both of them correspond to the onset of the second collapse.}\label{onset}%
\end{figure*}

In Figure~\ref{onset}, the profiles of $15M_{\odot}$ star and $40M_{\odot}$ star at the onset of the second collapse are compared. We can recognize that their profiles are similar for the innermost $0.5M_{\odot}$, where the phase transition to quark matter occurs. For the outer hadronic region, the $40M_{\odot}$ star has higher temperature and entropy than $15M_{\odot}$ star. Therefore, the core of $40M_{\odot}$ star ($1.68M_{\odot}$ in baryonic mass) is larger than that of $15M_{\odot}$ star ($1.44M_{\odot}$ in baryonic mass). After the second collapse, whole of the core converts to quark matter. While the thermal pressure is inefficient in the mixed phase, it is restored in the pure quark phase again. In particular, the collapse of $15M_{\odot}$ star, whose core is less massive than the maximum masses for $B=90$~fm$^{-3}$, bounces. On the other hand, a black hole is formed by the second collapse of $40M_{\odot}$ star. Note that, masses shown in Figure~\ref{rmass} are gravitational masses, which are different from baryonic masses in general. In fact, the maximum-mass star for $B=90$~fm$^{-3}$ is $1.54M_{\odot}$ in gravitational mass and $1.81M_{\odot}$ in baryonic mass. Thus, we can not predict the fate of the second collapse from a simple comparison of masses. Nevertheless, it is clear that whether the second core bounce occurs or not depends on the core mass.

\citet{sage09} and \citet{fish11} reported that the burst emission of electron type anti-neutrino accompanies the second bounce. Due to the stalled first shock, the shocked matter is deleptonized and becomes neutron-rich. When the second shock passes through the deleptonized region, the creation of electron-positron pairs occurs. Since this region is neutron-rich, positrons are well captured and copious electron type anti-neutrinos are produced. On the other hand, the electron capture is inadequate. The created electron-positron pairs also interact via the pair annihilation contributing to all neutrino species. Therefore, the amount of emitted neutrinos has a hierarchy $N_{\bar \nu_e} > N_{\nu_e} \sim N_{\nu_x}$, where $\nu_x \equiv \nu_\mu = \bar \nu_\mu = \nu_\tau = \bar \nu_\tau$. In this paper, the light curve of neutrino signal is not shown because we do not follow the long term evolution after the second bounce. However, in our computations, we confirm this hierarchy for the neutrinos produced by the second shock. These neutrinos would propagate outward and we would observe the burst emission of electron type anti-neutrino on Earth.

\section{Summary}

We have performed a series of core-collapse simulations taking into account the hadron-quark phase transition within the framework proposed by \citet{self08}. For the quark phase, we have used the MIT bag model and examined three cases for the bag constant, $B=90$, 150 and 250~MeV~fm$^{-3}$. Firstly, we have found that the collapse of $40M_{\odot}$ star results in the black hole formation for the all cases investigated and the phase transition hastens it. In particular, for the case with $B=90$~MeV~fm$^{-3}$, the shortening of the interval between the first bounce and black hole formation is drastic because quarks exist even at the early stage. Secondly, we have confirmed that the second collapse of $15M_{\odot}$ star bounces back and leads to explosion for the case with $B=90$~MeV~fm$^{-3}$. While this scenario had already been suggested by \citet{sage09}, we have shown it independently in this study for the different stellar model.

\begin{acknowledgements}
In this work, numerical computations were partially performed on the supercomputers at Research Center for Nuclear Physics (RCNP) in Osaka University, Center for Computational Astrophysics (CfCA) in the National Astronomical Observatory of Japan (NAOJ), Yukawa Institute for Theoretical Physics (YITP) in Kyoto University, Japan Atomic Energy Agency (JAEA), High Energy Accelerator Research Organization (KEK) and The University of Tokyo. This work was partially supported by Grants-in-Aids for the Scientific Research (Nos.~22540296, 23840038, 24244036) and Scientific Research on Innovative Areas (Nos.~20105004, 24105008) from the Ministry of Education, Culture, Sports, Science and Technology (MEXT) in Japan.
\end{acknowledgements}

\bibliographystyle{aa}
\bibliography{note.bbl}

\end{document}